\documentclass[aps,prl,twocolumn,preprintnumbers,amsmath,amssymb,superscriptaddress]{revtex4-1}
\pdfoutput=1

\usepackage{graphicx}
\usepackage{bm}
\usepackage[usenames,dvipsnames]{color}
\usepackage{hyperref}
\usepackage[tight]{subfigure}
\usepackage{verbatim}
\usepackage{units}
\usepackage{multirow}

\usepackage{t1enc}
\usepackage[polish,english]{babel}
\usepackage[cp1250]{inputenc}
\usepackage{polski}
\usepackage[T1]{fontenc}
\renewcommand{\vec}[1]{\mathbf{#1}}

\newcommand{\vecG}[1]{\bm{#1}}

\newcommand{\bra}[1]{\langle #1|}
\newcommand{\ket}[1]{|#1\rangle}

\usepackage[normalem]{ulem}

\begin{document}

\title{Effects of  Transverse Magnetic Anisotropy on  Current-Induced Spin Switching}

\author{Maciej Misiorny}
 \email{misiorny@amu.edu.pl}
\affiliation{Peter Gr{\"u}nberg Institut PGI-2, Forschungszentrum J{\"u}lich, 52425 J{\"u}lich,  Germany}
\affiliation{JARA\,--\,Fundamentals of Future Information Technology, 52425 J{\"u}lich,  Germany}
\affiliation{Faculty of Physics, Adam Mickiewicz University, Umultowska 85,  61-614 Pozna\'{n}, Poland}

\author{J\'{o}zef Barna\'{s}}
\affiliation{Faculty of Physics, Adam Mickiewicz University, Umultowska 85, 61-614 Pozna\'{n}, Poland}
\affiliation{Institute of Molecular Physics, Polish Academy of Sciences, Smoluchowskiego 17, 60-179 Pozna\'{n}, Poland}


\begin{abstract}
Spin-polarized transport through bistable magnetic adatoms or single-molecule magnets (SMMs), which  exhibit both uniaxial and transverse magnetic anisotropy, is considered theoretically. The main focus is on the impact  of transverse anisotropy on transport characteristics  and the adatom's/SMM's spin. In particular, we analyze the role of quantum tunneling of magnetization (QTM) in the mechanism of the current-induced spin switching, and show that the QTM phenomenon becomes revealed as resonant peaks in the average values of the molecule's spin and  in the charge current. These features  appear at some resonant fields and are observable when at least one of the electrodes is ferromagnetic. We also show that the conductance generally depends on the relative orientation of the average adatom's/SMM's spin and electrode's magnetic moment. This spin-valve like  magnetoresistance effect can be used to control spin switching of the adatom's/SMM's spin.

\end{abstract}

\pacs{72.25.-b,75.50.Xx,85.75.-d}


\maketitle


Experiments on electronic transport through individual atoms/molecules are at the forefront of the search for  novel nanoelectronics and information processing technologies. In this context, very prospective are magnetic atoms~\cite{Heinrich_Science306/2004,*Meier_Science320/2008,*Wiesendanger_Rev.Mod.Phys.81/2009,Hirjibehedin_Science317/2007,*Serrate_NatureNanotech.5/2010,Loth_NaturePhys.6/2010} and single-molecule magnets (SMMs)~\cite{Gatteschi_book,Heersche_Phys.Rev.Lett.96/2006,Zyazin_NanoLett.10/2010,*Burzuri_Phys.Rev.Lett.109/2012,Parks_Science328/2010,Vincent_Nature488/2012}  with a large spin $S\!\!>\!\!\frac{1}{2}$. If properly deposited onto a substrate, these quantum systems can acquire (in the case of atoms)~\cite{Brune_Surf.Sci.603/2009}  or retain (in the case of SMMs)~\cite{Mannini_NatureMater.8/2009,*Mannini_Adv.Mater.21/2009,*Kahle_NanoLett.12/2011} their intrinsic magnetic anisotropy -- a property responsible for  magnetic bistability.
Especially attracting is the idea of incorporating magnetic adatoms/SMMs into spintronic devices~\cite{Bogani_NatureMater.7/2008}, with the objective to use spin-polarized currents for manipulation of their magnetic moments~\cite{Misiorny_Phys.Rev.B75/2007,Timm_Phys.Rev.B73/2006,*Misiorny_Phys.Rev.B79/2009,*Delgado_Phys.Rev.Lett.104/2010,
*Delgado_Phys.Rev.B82/2010,*Fransson_Phys.Rev.B81/2010,*Sothmann_Phys.Rev.B82/2010,*Bode_Phys.Rev.B85/2012}. Actually, the feasibility of this concept has already been experimentally proven for Mn adatoms~\cite{Loth_NaturePhys.6/2010}. One of the key conditions for successful applications is a sufficiently large uniaxial magnetic anisotropy constant $D$. Therefore, some efforts have been undertaken in order to synthesize new molecules with large $D$ or to find other ways of anisotropy enhancement. It has been also demonstrated that magnetic anisotropy of an adatom/SMM can be systematically tuned, albeit in a limited range, e.g., by the environment adjustment~\cite{Otte_NaturePhys.4/2008}, an external electric field~\cite{Zyazin_NanoLett.10/2010,*Burzuri_Phys.Rev.Lett.109/2012}, or mechanical stretching  of a molecule~\cite{Parks_Science328/2010}.

Apart from the \emph{uniaxial} magnetic anisotropy underlying the magnetic bistability, adatoms and SMMs usually  possess also the \emph{transverse} component of the anisotropy~\cite{Gatteschi_book}. If the latter component is sufficiently large, it may lead to additional quantum effects, like oscillations due to the geometric Berry phase~\cite{Wernsdorfer_Science284/1999,*Leuenberger_Phys.Rev.Lett.97/2006} or quantum tunneling of magnetization (QTM)~\cite{Sessoli_Nature365/1993,*Thomas_Nature383/1996,*Mannini_Nature468/2010,Gatteschi_Angew.Chem.Int.Ed.42/2003}.
Although the role of QTM in electronic transport has been studied extensively for normal electrodes~\cite{Kim_Phys.Rev.Lett.92/2004,Romeike_Phys.Rev.Lett.96/2006,*Romeike_Phys.Rev.Lett.96/2006_196805,Gonzalez_Phys.Rev.Lett.98/2007,*Gonzalez_Phys.Rev.B78/2008}, much less is known how it affects the spin-polarized transport~\cite{Wang_Phys.Rev.B85/2012}. Since QTM allows for the underbarrier transitions between the states on the opposite sides of the energy barrier, it may serve as an additional dephasing mechanism, and thus impede the control of spin state by spin-polarized currents.

In this Letter we address the mechanism of current-induced spin switching in the presence of transverse anisotropy. We show that the conductance reveals peaks
at voltages where the thermal
transition rates
directly between degenerate  states of lowest energy are equal to the rate of transitions to the first excited state. Moreover, the transverse anisotropy significantly modifies the current-induced spin switching at some resonant fields, where the QTM phenomenon leads to resonant peaks in the field dependence of the average value of spin and charge current. These resonances are well pronounced in the field dependence of the derivative of current with respect to the magnetic field, and appear only when at least one electrode is ferromagnetic.
We  also show that the conductance generally depends on the relative orientation of the magnetic moments of the electrode and adatom/SMM. This effect follows from the interference of direct and indirect spin-conserving tunneling processes. In addition, a significant bias reversal asymmetry appears then in the transport characteristics.

\begin{figure}[t]
  \includegraphics[width=0.99\columnwidth]{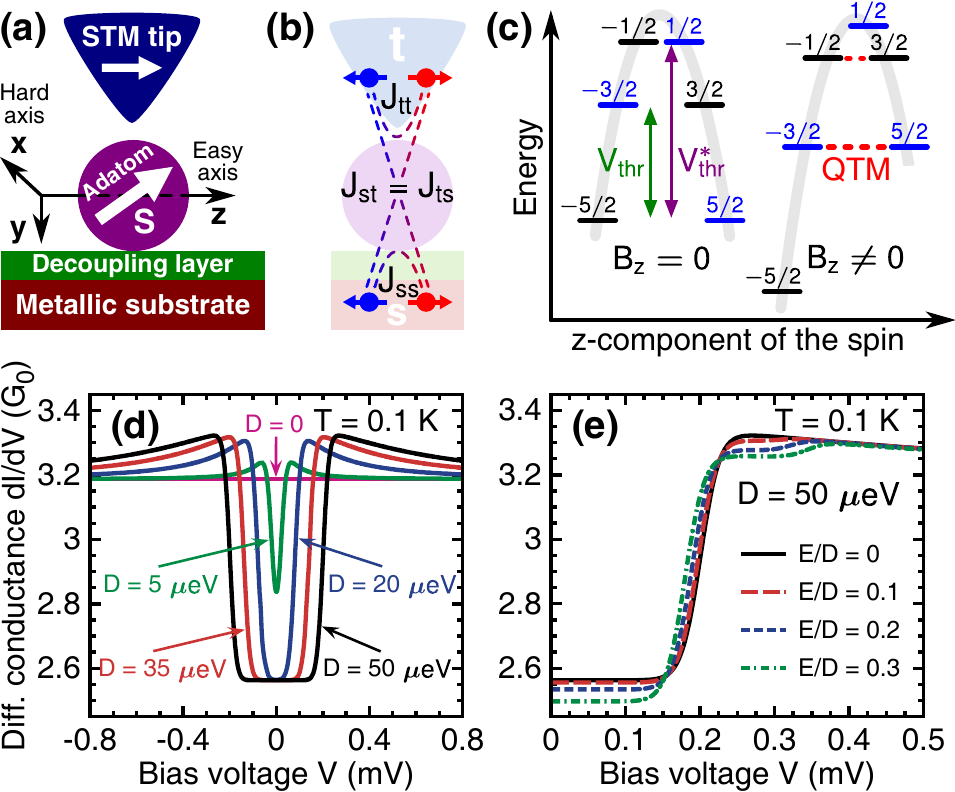}
  \caption{\label{Fig:1} (Color online) (a) Schematic depiction of the system under consideration. (b) Examples of inelastic electron tunneling processes due to scattering of electron spin on the adatom's/SMM's spin. (c) Energy spectrum of the adatom/SMM for $S\!\!=\!\!5/2$ in the absence (left) and presence (right) of an external magnetic field along the easy axis, $\vec{B}=(0,0,B_z)$. Bottom panel shows the differential conductance  as a function of bias voltage in the case of \emph{nonmagnetic} electrodes ($P_t=P_s=0$) for: (d) selected values of the uniaxial magnetic anisotropy constants $D$ and absence of transverse anisotropy, $E=0$; (e) several values of $E$ for a given $D$. Remaining parameters: $T_\textrm{d}=0.1$ eV, $\rho^t=\rho^s=0.5$ eV$^{-1}$ [thus $G_0\approx0.025$ $\frac{2e^2}{h}$], $\alpha=1$ and $2\eta_t=\eta_s=1$.
 }
\vspace*{-5pt}
\end{figure}

\emph{Model.---}Key features of
magnetic adatoms and SMMs are captured by the \emph{giant-spin Hamiltonian}~\cite{Gatteschi_book},
    \begin{equation}\label{Eq:H_0}
\vspace*{-2pt}
    \mathcal{H_S}=
    -DS_z^2
    +
     \frac{E}{2}(S_+^2+S_-^2)
    +
    \vec{S}\cdot\vec{B},
\vspace*{-2pt}
    \end{equation}
where the first and second terms stand for the \emph{uniaxial} and \emph{transverse}  magnetic anisotropy, respectively, while the last term represents the Zeeman interaction, with $\vec{B}\!=\!(B_x,B_y,B_z)$ denoting an external magnetic field measured in energy units. Since we are interested here  in systems with an energy barrier for spin switching, we assume $D\!>\!0$. Without losing generality, we also assume positive perpendicular anisotropy constant, $E>0$, and $0\!\leqslant\! E/D\! \leqslant\! \tfrac{1}{3}$~\cite{Gatteschi_book}. When $E\!\neq\!0$, each of the $2S+1$ eigenstates $\ket{\chi}$ of the Hamiltonian~(\ref{Eq:H_0}), $\mathcal{H_S}\ket{\chi}=E_\chi\ket{\chi}$, is  a linear combination of the eigenstates $\ket{m}$ of the $S_z$ component. We label the states $\ket{\chi}$ with a subscript $m$, $\ket{\chi}\!\to\! \ket{\chi_{m}}$, which [as well as the numbers in Fig.~\ref{Fig:1}(c)] corresponds to the  $S_z$ component of highest weight in the state $\ket{\chi_{m}}$, i.e. $\ket{\chi_m}\!\equiv\!\ket{m}$ for $E\!\to\!0$. When $\vec{B}\!=\!(0,0,B_z)$, the eigenstates $\ket{\chi_m}$
can be written as
    $
    \ket{\chi_m}\!=\!\sum_{k=0,\pm2}\left<m+k|\chi_m\right>\ket{m+k}
    $
for $m=-S,\ldots,S$, where $\left<m|\chi_m\right>$ is the amplitude of the state $\ket{m}$ in the eigenstate $\ket{\chi_{m}}$.  Thus, the transverse anisotropy leads to mixing of the states $\ket{m}$, and therefore enables the QTM~\cite{Gatteschi_Angew.Chem.Int.Ed.42/2003}. In the following, the index $m$ shall be used only when necessary to avoid any confusion.

We consider an experimental configuration including tip of the scanning tunneling microscope (STM), and a substrate [which plays the role of second electrode] at which the adatom/SMM is deposited, see Fig.~\ref{Fig:1}(a). Both, STM tip and substrate are characterized by noninteracting itinerant electrons,
    $
     \mathcal{H}_\textrm{el}\!=\!
     \sum_{q\vec{k}\sigma}
     \varepsilon_{\vec{k}\sigma}^q
     a_{\vec{k}\sigma}^{q\dag} a_{\vec{k}\sigma}^q
    $
($q\!=\!t$ for the STM tip, and $q\!=\!s$ for the substrate), with the
energy dispersion $\varepsilon_{\vec{k}\sigma}^q$, and $a_{\vec{k}\sigma}^{q\dag}$ ($a_{\vec{k}\sigma}^{q}$) being the relevant creation (annihilation) operators ($\vec{k}$ is a wave vector, and $\sigma$ is the electron spin index). In general, both electrodes can be magnetic, with spin-dependent density of states (DOS) $\rho_\sigma^q$ at the Fermi level. By introducing the \emph{spin polarization coefficient}, $P_q\!=\!(\rho_\uparrow^q-\rho_\downarrow^q)/(\rho_\uparrow^q+\rho_\downarrow^q)$, the DOS can be parameterized as: $\rho_{\uparrow(\downarrow)}^q=\frac{\rho^q}{2}(1\pm P_q)$  with $\rho^q=\rho_\uparrow^q+\rho_\downarrow^q$.

Electron tunneling processes  in the STM geometry are modeled by the Appelbaum Hamiltonian~\cite{Appelbaum_Phys.Rev.154/1967,*Appelbaum_Phys.Rev.Lett.17/1966,Nussinov_Phys.Rev.B68/2003,*Fransson_NanoLett.9/2009,Kim_Phys.Rev.Lett.92/2004,Misiorny_Phys.Rev.B75/2007},
    \begin{equation}\label{Eq:H_tun}
\vspace*{-2pt}
    \mathcal{H}_\textrm{T}
    =\!
    \sum_{q\vec{k}\vec{k}'\alpha}
    \!\!
    \Big\{
    T_\textrm{d}\,
    a_{\vec{k}\alpha}^{q\dag}
    a_{\vec{k}'\alpha}^{\bar{q}}
    +
    \sum_{q^\prime\beta}
    \!
    J_{qq^\prime}\,
    \vecG{\sigma}_{\alpha\beta}\cdot\vec{S}\:
    a_{\vec{k}\alpha}^{q\dag}
    a_{\vec{k}'\beta}^{q^\prime}
    \Big\},
\vspace*{-2pt}
    \end{equation}
with $\bar{q}$ to be understood as $\bar{s}\equiv t$ and $\bar{t}\equiv s$.  Electrons can tunnel either directly between the two electrodes [the first term  Eq.~(\ref{Eq:H_tun})], or during the tunneling event they can interact with the adatom/SMM \emph{via} exchange coupling [the second term in Eq.~(\ref{Eq:H_tun})]. The former processes are described by the tunneling parameter $T_\textrm{d}$, whereas the latter ones by the exchange parameter $J_{qq^\prime}$, see Fig.~\ref{Fig:1}(b). Both $T_\textrm{d}$ and $J_{qq^\prime}$ are assumed to be real, isotropic, and independent of energy and the electrodes' spin-polarization. In the following discussion, we write $J_{qq^\prime}=J\eta_q\eta_{q^\prime}$, with $\eta_q$ denoting the dimensionless scaling factor of the coupling between the adatom/SMM and the $q$th electrode (we fix $\eta_s=1$).  We also  relate the parameters $J$ and $T_\textrm{d}$ as $J\equiv\alpha T_\textrm{d}$. Thus, $T_\textrm{d}$ will serve as the key, experimentally relevant parameter~\cite{Nussinov_Phys.Rev.B68/2003,*Fransson_NanoLett.9/2009}, and $\alpha$ describes relation between the direct tunneling processes and those involving spin scattering of conduction electrons. For simplicity, electrodes' magnetic moments are assumed to be collinear with the easy axis of the adatom/SMM.


\emph{Method.---}In the weak coupling regime, transport characteristics can be derived using the approach based on a  master equation. The charge current ($e\!\!>\!\!0$) flowing between the STM tip and the substrate is then given by
    $
    I=e\sum_{\chi\chi^\prime} \mathcal{P}_\chi
    \big\{
    \gamma_{\chi\chi^\prime}^{ts}-\gamma_{\chi\chi^\prime}^{st}\!
    \big\},
    $
where $P_{\chi}$ is the probability of finding the SMM/adatom  in the spin state $\ket{\chi}$ ($\equiv\ket{\chi_m}$), and $\gamma_{\chi\chi^\prime}^{qq^\prime}$ stands for the transition rate between the states $\ket{\chi}$ and $\ket{\chi^\prime}$ ($\equiv\ket{\chi_{m^\prime}}$) associated with electron tunneling between the electrodes $q$ and $q^\prime$.

For the sake of analytical clarity, we decompose  the total current into two parts, $I\!=\!I_\textrm{el}\!+\!I_\textrm{in}$,  where  $I_\textrm{el}$ represents  the contribution due to \emph{elastic} electron tunneling processes [the  spin state $\ket{\chi}$ remains unchanged] and $I_\textrm{in}$ is the \emph{inelastic} term [with transitions between different states $\ket{\chi}$ and $\ket{\chi^\prime}$]. In the second order approximation with respect to the electrode-SMM/adatom coupling, these two  current components take the form
    \begin{gather}\label{Eq:I_el}
    \vspace*{-2pt}
    \hspace*{-7pt}
    \frac{I_\textrm{el}}{G_0}=
    \big\{
    1
    +
    P_tP_s
    +
    2\widetilde{\alpha}
    \langle S_z\rangle
    (P_t+P_s)
    \big\}
    V
    +
    \sum_{\chi}
    \mathcal{P}_\chi
    \mathcal{V}_{\chi\chi}
    ,
            \\\label{Eq:I_inel}
    \frac{I_\textrm{in}}{G_0}
    =
    \sum_\chi
    \sum_{\chi^\prime(\neq\chi )}
    \!
    \mathcal{P}_\chi
    \mathcal{V}_{\chi^\prime\chi}
    ,
    \vspace*{-2pt}
    \end{gather}
where
    \begin{align}
    \hspace*{-3pt}
    \mathcal{V}_{\chi^\prime\chi}&=
    \tfrac{\widetilde{\alpha}^2}{e}
    \Bigg\{
    \Big[
    \big|\mathbb{S}_{\chi^\prime\chi}^z\big|^2
     +
    \tfrac{1}{2}
    \sum_{\kappa=\pm}
    \big|\mathbb{S}_{\chi^\prime\chi}^\kappa\big|^2
    \Big]
    Z_-\!\big(\Delta^{\chi\chi^\prime}_{eV}\big)
                \nonumber\\[-2.5pt]
    &
    +
    P_tP_s
    \Big[
    \big|\mathbb{S}_{\chi^\prime\chi}^z\big|^2
     -
    \tfrac{1}{2}
    \sum_{\kappa=\pm}
    \big|\mathbb{S}_{\chi^\prime\chi}^\kappa\big|^2
    \Big]
    Z_-\!\big(\Delta^{\chi\chi^\prime}_{eV}\big)
                \nonumber\\[-10pt]
   &\hspace*{27pt} +
    \tfrac{1}{2}
    (P_t-P_s)
    \sum_{\kappa=\pm}
    \kappa
    \big|\mathbb{S}_{\chi^\prime\chi}^\kappa\big|^2
     Z_+\!\big(\Delta^{\chi\chi^\prime}_{eV}\big)
    \Bigg\}.
    \end{align}
In the equations above,  $G_0\equiv\frac{\pi e^2}{\hbar} \rho^t\rho^s |T_\textrm{d}|^2$, $\widetilde{\alpha}\equiv\alpha\eta_t\eta_s$ and $\mathbb{S}_{\chi^\prime\chi}^\kappa\equiv\bra{\chi^\prime}S_\kappa\ket{\chi}$ for $\kappa=z,\pm$. Accordingly, $\langle S_z\rangle\!=\!\sum_\chi \mathcal{P}_\chi \mathbb{S}_{\chi\chi}^z$. In addition,
$\Delta^{\chi\chi^\prime}_{eV~}\!\!\equiv\! E_\chi-E_{\chi^\prime} + eV$, where $eV=\mu_t-\mu_s$  stands for the difference in electrochemical potentials of the tip ($\mu_t$) and substrate ($\mu_s$), while
    $
    Z_\pm\!\big(\Delta^{\chi\chi^\prime}_{eV}\big)
    \equiv
    \zeta\big(\Delta^{\chi\chi^\prime}_{eV}\big)
    \pm
    \zeta\big(\Delta^{\chi\chi^\prime}_{-eV}\big)
    $
with $\zeta(x)\equiv x/\big\{1-\exp(-x/\{k_\textrm{B}T\})\big\}$ and $T$ denoting temperature.

In order to evaluate the current from Eqs.~(\ref{Eq:I_el}) and (\ref{Eq:I_inel}), we need the probabilities $P_\chi$. These can be obtained from the set of stationary master equations, $\forall_\chi$:
    $
    \sum_{\chi^\prime}
    \!
    \Big\{\!
    P_{\chi^\prime} \gamma_{\chi^\prime\chi}
    -
    P_{\chi} \gamma_{\chi\chi^\prime}
    \!\Big\}
    =0,
    $
where the golden rule transition rates $\gamma_{\chi\chi^\prime}=\sum_{qq^\prime}\gamma_{\chi\chi^\prime}^{qq^\prime}$ for $\chi\neq\chi^\prime$ are given by
    \begin{multline}
     \hspace*{-8pt}
    \gamma_{\chi\chi^\prime}^{qq^\prime}  =   \frac{2\pi}{\hbar} |T_\textrm{d}|^2
    (\alpha\eta_q\eta_{q^\prime})^2
    \zeta\big(\Delta^{\chi\chi^\prime}_{\mu_{q}-\mu_{q^\prime}}\big)
                \\
    \hspace*{8pt}
   \times
   \!
    \Big[
    \sum_{\sigma} \rho_\sigma^q\rho_\sigma^{q^\prime}
    \big|\mathbb{S}_{\chi^\prime\chi}^z\big|^2
    \!+\!
    \rho_\uparrow^q\rho_\downarrow^{q^\prime}
    \big|\mathbb{S}_{\chi^\prime\chi}^+\big|^2
    \!+\!
    \rho_\downarrow^q\rho_\uparrow^{q^\prime}
    \big|\mathbb{S}_{\chi^\prime\chi}^-\big|^2
    \Big]
   .
    \end{multline}
%


\emph{Nonmagnetic electrodes.---}Consider a model system of spin $S\!\!=\!\!5/2$, connected to nonmagnetic tip and substrate, and characterized by typical parameters observed in experiments, see the caption of Fig.~\ref{Fig:1}. For vanishingly small  transverse magnetic anisotropy ($E=0$) and $|\vec{B}|=0$, the Hamiltonian~(\ref{Eq:H_0}) is diagonal in  the basis of the eigenstates of $S_z$. As a result, Eqs.~(\ref{Eq:I_el}) and (\ref{Eq:I_inel}) simplify significantly~\cite{Misiorny_Phys.Rev.B75/2007}:
    $
    I_\textrm{el}^{E=0}/G_0
    \!=\!
    \big\{
    1+
    \widetilde{\alpha}^2
    \langle S_z^2\rangle
    \big\}
    V
    $
and
    $
    I_\textrm{in}^{E=0}/G_0
    \!=\!
    \frac{\widetilde{\alpha}^2}{2e}
    \sum_m
    \mathcal{P}_m\!
    \sum_{\lambda=\pm1}\!
    \big[
    A_\lambda(m)
    \big]^2
    Z_-\!\big(\Delta^{m,m+\lambda}_{eV}\big)
    $
with $A_\lambda(m)=[{S(S+1)-m(m+\lambda)}]^{1/2}$.
For $D>0$ and low $T$, the system occupies then with equal probabilities each of the metastable ground states $\ket{\!\pm S}$. When  bias voltage $|V|$ increases, initially only the elastic tunneling processes contribute to transport, i.e. $I_\textrm{el}^{E=0}\neq0$ and $I_\textrm{in}^{E=0}\approx0$. When $|V|$ becomes of the order of the threshold value $V_\textrm{thr}=D(2S-1)=4D$ (for $S=5/2$), see Fig.~\ref{Fig:1}(c),  the inelastic processes become activated  and  $I_\textrm{in}^{E=0}\neq0$, which appears as a characteristic step in the differential conductance, Fig.~\ref{Fig:1}(d,e). This feature is typical of the systems exhibiting \emph{easy-axis} magnetic anisotropy~\cite{Hirjibehedin_Science317/2007,Loth_NaturePhys.6/2010}.

 For  $E\!>\!0$
 and a half-integer spin $S$, the eigenstates $\ket{\chi_m}$ are twofold degenerate (Kramers' doublets) and form two uncoupled sets $\big\{\ket{\chi_{\pm S\mp2k}}\big\}_{k=0,1,\ldots, S-1/2}$~\cite{Romeike_Phys.Rev.Lett.96/2006}, schematically distinguished by  different colors in Fig.~\ref{Fig:1}(c).  Mixing of the states corresponding to different values of $m$ is also revealed  in the expressions for current,
    $
    I_\textrm{el}^{E\neq0}/G_0
    \!=\!
    \big\{
    1+
    \widetilde{\alpha}^2
    \sum_{k}
    \mathcal{P}_{\chi_{k}}
    \big|\sum_m |\!\left< m|\chi_{k}\right>\!|^2m\big|^2
    \big\}
    V
    $
and
    $
    I_\textrm{in}^{E\neq0}/G_0
    \!=\!
    \frac{\widetilde{\alpha}^2}{e}
    \sum_{k}
    \sum_{l(\neq k)}
    \mathcal{P}_{\chi_k}
    \big\{
    \big|\sum_m \left< \chi_l|m\right>\!\left<m|\chi_k\right>  m\big|^2
    +
    \frac{1}{2}
    \sum_{\lambda=\pm1}
    \big|\sum_m \left< \chi_l|m+\lambda\right>\!\left<m|\chi_k\right> A_\lambda(m)\big|^2
    \big\}
    Z_-\!\big(\Delta^{\chi_k\chi_l}_{eV}\big).
    $

The transverse anisotropy manifests as several new features in transport characteristics. First, for $|V|\!\lesssim\!V_\textrm{thr}$ it leads to reduction of~$I_\textrm{el}$ [cf. the second terms of $I_\textrm{el}^{E=0}$ and $I_\textrm{el}^{E\neq0}$ given above], which appears as a decreased value of $\textrm{d}I/\textrm{d}V$, Fig.~\ref{Fig:1}(e). Second, the transverse anisotropy effectively gives rise  to a small reduction of~$V_\textrm{thr}$.
Actually, at equilibrium, $V\approx0$, the spin can
directly oscillate  between the two ground states $\ket{\chi_{\pm S}}$ as $\gamma_{\chi_{-S}\chi_S}=\gamma_{\chi_S\chi_{-S}}\!\propto\!k_\textrm{B}T$~\cite{SuppInfo}, which is in contrast to $\gamma_{-SS}=\gamma_{S-S}=0$ for $E=0$. Such
`underbarrier'
oscillations will dominate until they are surpassed by transitions to the first excited spin dublet, which occurs at $|V|\approx V_\textrm{thr}(E)$. It's worth emphasizing that the effect stems entirely from thermal fluctuations, so that the Kramers' degeneracy is not affected. Finally, for significantly large transverse magnetic anisotropy the second step in the differential conductance appears at $V_\textrm{thr}^\ast\approx4D(S-1)=6D$ (for $S=5/2$), when  direct spin excitations to the second excited Kramers' doublet become possible, Fig.~\ref{Fig:1}(c). We note, however, that these features are clearly distinguishable only if $E\gg T$.


\begin{figure}[t]
  \includegraphics[width=0.99\columnwidth]{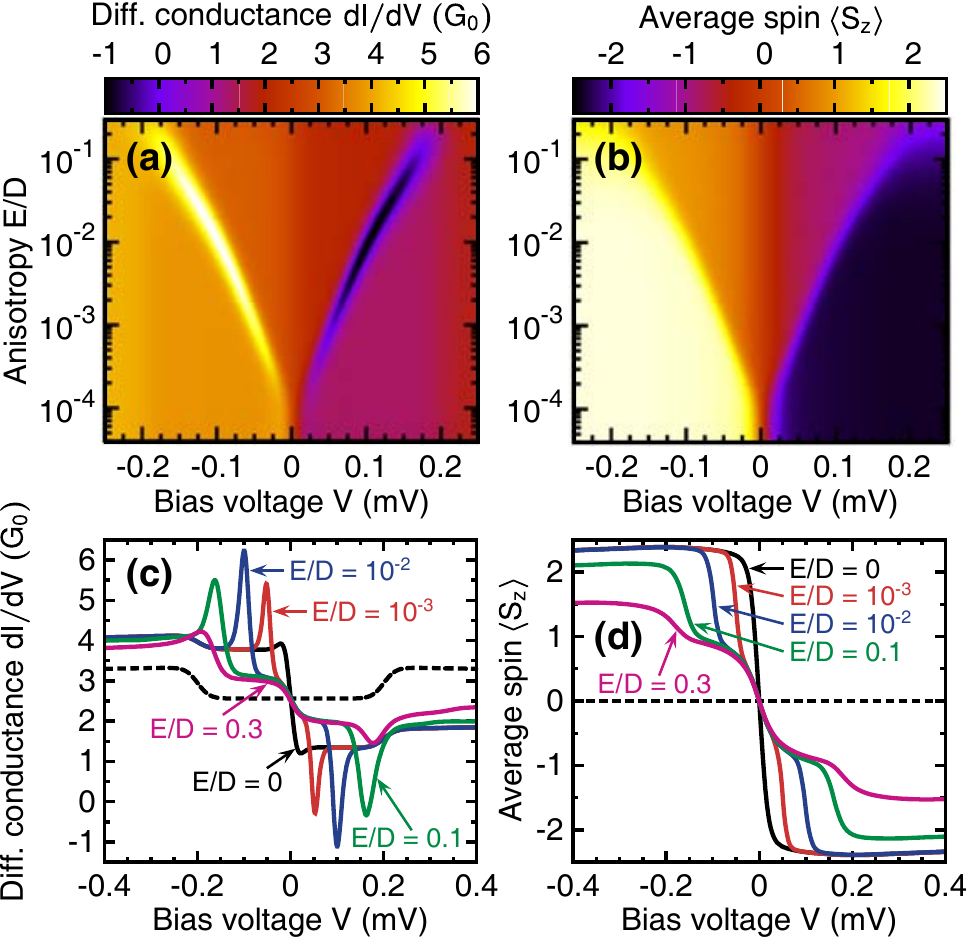}
  \caption{\label{Fig:2} (Color online) Dependence of differential conductance (a) and the average value of the  spin's $z$th component (b) on bias voltage and transverse magnetic anisotropy for $D=50$ $\mu$eV and $P_t=0.5$. Solid lines in (c) and (d) represent cross-sections of (a) and (b), respectively,  for selected values of $E$, while the dashed line corresponds to $E/D=0$ and $P_t=0$. Other parameters as in Fig.~\ref{Fig:1}.
 }
 \vspace*{-5pt}
\end{figure}

\emph{Current induced magnetic switching.---}In order to control the adatom/SMM's spin state by  a spin-polarized current, at least one of electrodes has to be magnetic~\cite{Misiorny_Phys.Rev.B75/2007,Timm_Phys.Rev.B73/2006}. Here, we choose it to be the tip, $P_t\neq0$ and $P_s=0$. Consider first the case of $E=0$. When $|V|< T$, the system's spin can still fluctuate
indirectly between the states $\ket{\pm S}$
as a result of  stepwise `overbarrier' transitions. However, due to imbalance of the spin-up and spin-down electron tunneling processes, the systems's spin becomes locked in one of the ground states $\ket{\pm S}$ [depending on the bias polarization] for $|V|\gtrsim T$, Fig~\ref{Fig:2}(d). In consequence, the conductance is  determined in the GMR-fashion  by the relative orientation of the tip's magnetization and the adatom's/SMM's spin, $\textrm{d}I/\textrm{d}V\propto2\widetilde{\alpha} P_t\langle S_z\rangle $, Fig~\ref{Fig:2}(c). This behavior stems from
interference of direct tunneling and spin conserving part of tunneling associated with exchange interaction between tunneling electrons and magnetic core of an adatom/SMM, and also leads to the asymmetry in conductance with respect to bias reversal, see Fig.~\ref{Fig:2}(c).
Inelastic transport processes activated at $|V|\!\approx\!V_\textrm{thr}$  enhance conductance [see the curve for $E/D=0$ in Fig~\ref{Fig:2}(c)], but slightly reduce the absolute value of the average spin $S_z$ component [see the curve for $E/D=0$ in Fig~\ref{Fig:2}(d)].

The situation changes qualitatively for $E\neq0$. For low voltages, the spin  oscillates between $\ket{\chi_{\pm S}}$, like in the nonmagnetic case. However, due to the spin-dependence of tunneling processes $\gamma_{\chi_{-S}\chi_S} \neq \gamma_{\chi_S\chi_{-S}}$, and the spin generally resides longer in one of the two ground states~\cite{SuppInfo}. At some voltage,
transition rate to the excited state surpasses the transition rate between the
two ground states and the spin becomes locked in one of the two states $\ket{\chi_{\pm S}}$. This results in an additional peak  in $\textrm{d}I/\textrm{d}V$, Fig.~\ref{Fig:2}(c). Since the transition rate $\gamma_{\chi_{-S}\chi_S}$ increases with~$E$, position of this peak moves towards larger voltage
with augmenting $E$. Further rise
in voltage leads to saturation of the conductance, and the saturated value only weakly depends on $E$. It's worth of note, however, that the strong mixing of spin states for large transverse anisotropy prevents the system's spin from aligning along the easy axis.  Surprisingly, in the case of spin-polarized transport, the interplay of the underbarrier relaxation process introduced by  the transverse magnetic anisotropy imposes the voltage barrier for switching the system's spin. Such a behavior doesn't take place in systems exhibiting purely uniaxial magnetic anisotropy.


\begin{figure}[t]
  \includegraphics[width=0.99\columnwidth]{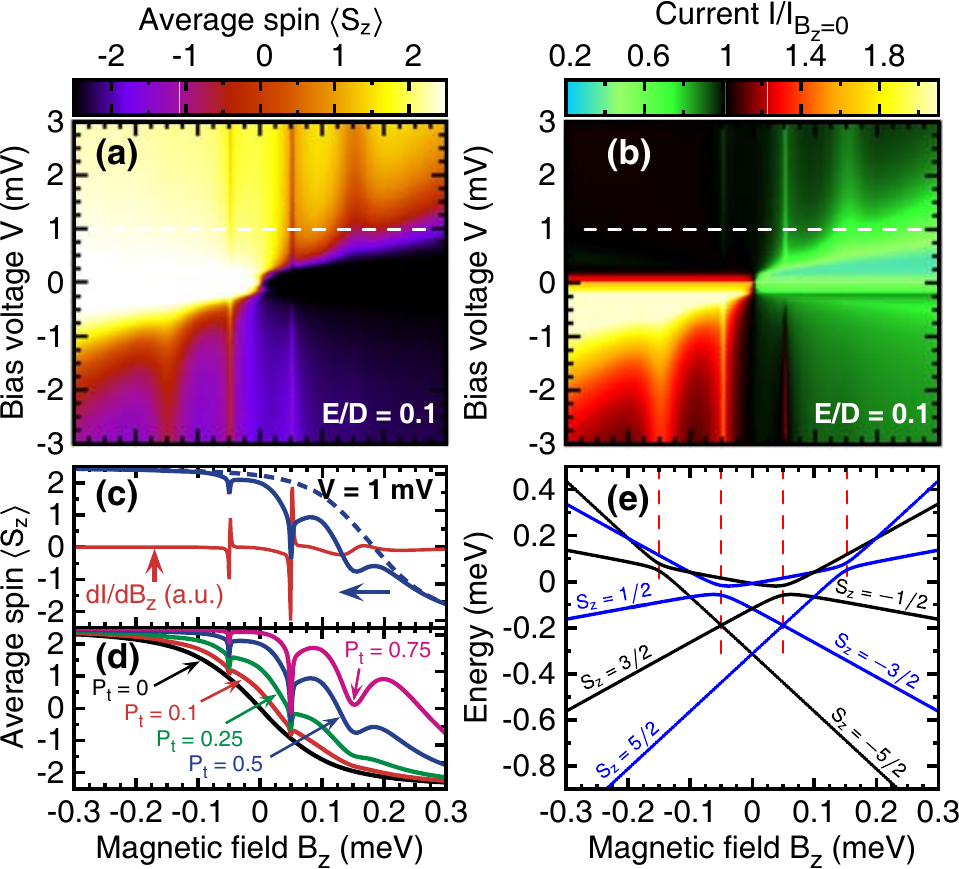}
  \caption{\label{Fig:3} (Color online) (a) The average value of the spin's $z$th component and (b) the charge current shown as functions of an external magnetic field applied along the system's easy axis, $B_z$,   and bias voltage for $D=50$ $\mu$eV, $E/D=0.1$ and $P_t=0.5$.  (c) Dark bold line represents a cross-section of (a) for indicated bias voltage [dashed  lines correspond here to the case of $E=0$], while bright bold line corresponds to $\textrm{d}I/\textrm{d}B_z$, shown here in arbitrary units. (d) Average spin for $V\!=\!1$ mV, $E/D=0.1$, and for several values of the tip polarization $P_t$. (e) Dependence of the system's energy spectrum on magnetic field $B_z$. Other parameters as in Fig.~\ref{Fig:1}.
 }
 \vspace*{-5pt}
\end{figure}

When  an external magnetic field is applied, one can achieve degeneration of different states belonging to either of the two decoupled manifolds, see
the right side of Fig.~\ref{Fig:1}(c). This degeneration, though, is not complete due to level repulsion, as shown in Fig.~\ref{Fig:3}(e). Anyway, at the fields corresponding to the dashed lines in Fig.~\ref{Fig:3}(e), quantum tunneling of magnetization takes place, which results in transitions between the two degenerate states. These transitions are clearly seen as resonant peaks in the average values of the molecules's spin, Fig.~\ref{Fig:3}(a,c,d), and current, Fig.~\ref{Fig:3}(b). The resonant character of QTM is even more evident in the field dependence of $\textrm{d}I/\textrm{d}B_z$, see Fig.~\ref{Fig:3}(c). Interestingly, the resonances due to QTM  can be observed only when (at least) one of the electrodes  is ferromagnetic, as follows from Fig.~\ref{Fig:3}(d).


\emph{Conclusions.---}
In this Letter we have considered the influence of transverse magnetic anisotropy on spin polarized transport through magnetic adatoms/molecules, and in particular
on the current-induced spin switching. First, we have demonstrated that mixing of states by transverse anisotropy leads to a decrease in
conductance in the elastic transport regime (low voltage regime) and to appearing of the peaks at voltages where the system's spin becomes locked in one of the two ground states. When an external magnetic field is applied, the phenomenon of quantum tunneling of magnetization, which occurs  at some resonant values of the magnetic field, results in  resonant peaks in the average value of the molecules's spin and in the charge current.
It is worth emphasizing, however,  that these effects  can be observed only when the tip (and/or substrate) is ferromagnetic. Thus, spin-polarized transport spectroscopy may prove a useful experimental tool for studying the phenomenon of QTM.

It has been  also shown that the conductance generally depends on the relative orientation of the average adatom's/SMM's spin and electrode's magnetic moment. This dependence stems from the interference of direct tunneling and spin conserving tunneling connected with exchange interaction between the electrons and adatom/SMM spin, and resembles the giant magnetoresistance effect in magnetic multilayers. It is also responsible for a significant asymmetry of transport characteristics with  respect to
the bias reversal and can  be used to control spin switching of the adatom's/SMM's spin.


\emph{Acknowledgemnts.---}M.M. acknowledges support from the Foundation for Polish Science and  the Alexander von Humboldt Foundation. The work has been supported by National Science Center in Poland as a research
project in years 2010-2013 and the Project No. DEC-2012/04/A/ST3/00372.


%

\end{document}